\documentclass[twocolumn,apj,appendixfloats,numberedappendix]{emulateapj-rtx4}

\usepackage{natbib}
\usepackage{graphicx}
\usepackage{amsmath,amssymb}
\usepackage{ulem}
\usepackage{enumitem}
\usepackage{txfonts}
\usepackage{epstopdf}

\newcommand{\vect}[1]{{\boldsymbol{#1}}}
\newcommand{\bnabla}{\boldsymbol{\nabla}}

\newcommand{\chb}[1]{{#1}}

\newcommand{\be}{\begin{equation}}
\newcommand{\ee}{\end{equation}}
\newcommand{\ba}{\begin{eqnarray}}
\newcommand{\ea}{\end{eqnarray}}

\begin{document}

\title{3D anisotropy of solar wind turbulence, tubes or ribbons?}

\author{Andrea~Verdini}
\affil{Universit\'a di Firenze, Dipartimento di Fisica e Astronomia, Firenze, Italy}
\author{Roland~Grappin}
\affil{LPP, Ecole Polytechnique, Palaiseau, France.}
\author{Olga~Alexandrova}
\affil{Lesia, Observatoire de Paris, Meudon, France}
\author{Sonny~Lion}
\affil{Lesia, Observatoire de Paris, Meudon, France}


\date{\today}

\begin{abstract}
We study the anisotropy with respect to the local magnetic field of turbulent \chb{magnetic} fluctuations at magnetofluid scales in the solar wind.
Previous measurements assumed \chb{the mean field as a local symmetry axis and obtained axisymmetric anisotropy}. However, results are probably contaminated by the wind expansion that introduces another \chb{symmetry axis}, namely the radial direction, \chb{along which measurements are taken}.
Recent numerical simulations show that while the expansion is strong, the principal fluctuations are in the plane perpendicular to the radial direction. 
Using this property, we separate 10 years of Wind spacecraft data into two subsets characterized  by strong and weak expansion and determine the corresponding turbulence anisotropy.
Under strong expansion, \chb{the small-scale} anisotropy is consistent with Goldreich \& Sridhar (1995) critical balance. \chb{As in previous works, when the radial symmetry axis is not eliminated}, the turbulent structures are field aligned tubes. 
Under weak expansion, we find 3D anisotropy predicted by Boldyrev (2006) model, that is, turbulent structures are ribbons and not tubes. However, the very basis of the Boldyrev phenomenology, namely a cross-helicity increasing at small scales, is not observed in the solar wind: the origin of the ribbon formation is unknown.
\end{abstract}
\maketitle

\section{Introduction}
The physics of the magnetohydrodynamic (MHD) turbulence in plasmas threaded by a large-scale mean magnetic field ($B_0$) is known to be anisotropic. 
Anisotropy can be measured with respect to $B_0$ (global anisotropy) or to the local mean field, $B_l$ (local anisotropy), where local is meant both in space and in scales \citep{Cho_Vishniac_2000}. Note that the latter shows up even in absence of a large-scale mean field.

Global anisotropy is caused by gradients developing preferentially in directions perpendicular to $B_0$ and the resulting spectrum extends in perpendicular wavenumbers \citep{Montgomery_Turner_1981,Shebalin_al_1983,Grappin_1986}. 
The tendency of MHD turbulence to become two-dimensional (2D) can be traced in the hierarchy of higher order moments: 
the cascade rate measured with third-order moments is itself anisotropic, causing the anisotropy observed in spectra \citep{Verdini_al_2015}, while the fourth-order moments have an explicit dependence \chb{on} the mean field which in turn causes anisotropy in the cascade rate \citep{Oughton_al_2013}.

Measurements in the solar wind confirmed that turbulence has a main 2D component \citep{Matthaeus_al_1990}, but further observations highligthed the presence of a second population, the so-called slab component with axis of symmetry given by the mean field \citep{Bieber_al_1996} or by the radial direction \citep{Saur_Bieber_1999}.
\chb{It has been recently shown that the 2D and field-aligned slab components emerge naturally in the formulation of nearly incompressible MHD equations \citep{Zank_al_2017}. However, It is not clear yet if they can coexist in MHD turbulence \citep{Ghosh_al_1998b,Ghosh_al_1998a} and if field-aligned wavectors can be maintained in the solar wind \citep{Volk_Aplers_1973,Grappin_al_1993}.
On the other hand, \citet{Verdini_Grappin_2016} provided an explanation for the detection of field-aligned slab and radially-aligned slab components that is based on expansion effects. Using numerical simulations that include the effect of solar wind expansion (Expanding Box Model, EBM, \citealt{Grappin_al_1993,Grappin_Velli_1996}), they showed that the radial axis of symmetry is an attractor for turbulence anisotropy. Radial anisotropy  prevails on the anisotropy due to the mean field (the other attractor) when field-aligned wavenumbers are excited at 0.2 AU. Thus, the observed field-aligned slab component might originate from the (wrong) assumption of axisymmetry around the mean field when measuring a truly radially-aligned slab component.}

The scenarios dealing with local anisotropy consider again the mean field (now, the local mean field, $B_l$) as the only symmetry axis, with gradients developing mainly perpendicularly to it. To partition the anisotropy with respect to the local reference frame, whose orientation depends on the scale and of the region \chb{on} interest, structure functions or wavelets are used instead of second order correlation or spectra (hereafter we will refer to structure function, $SF$ to describe the local anisotropy). 

Physically, in strong MHD turbulence the local anisotropy is supposed to be constrained by the so-called critical balance between the nonlinear cascade time and the linear transport time along field lines \citep{GS95}. 
If the cascade time is fast, i.e. equal to the eddy-turnover time,
the local structure function has a Kolmogorov-like spectral index $SF\sim\ell_\bot^{2/3}$, where $\ell_\bot$ is the increment perpendicular to the local mean field, while it has a steeper power-law scaling $SF_\|\sim\ell_\|$ for parallel increments $\ell_\|$ (hereafter GS phenomenology).
Critical balance implies $\ell_\|\sim\ell_\bot^{2/3}$: contrary to the global anisotropy, the local anisotropy is scale dependent and eddies at small scales become filaments (or tubes) that are aligned to the local mean field.

Is the mean field the only symmetry axis for local anisotropy? 
Measurements in the solar wind assuming axisymmetry around the local mean field obtained structure functions that scale as
$SF\sim\ell_\bot^{2/3}$ and $SF_\|\sim\ell_\|$ \citep{Horbury_al_2008, Podesta_2009, Luo_Wu_2010, Wicks_al_2010, Wicks_al_2011}, that is, a local anisotropy consistent with GS phenomenology.
However, recently the only available three-dimensional (3D) measurement shows that anisotropy changes qualitatively with scales \citep{Chen_al_2012}. Eddies are aligned to the displacement direction at large scales and become aligned to the mean field only at small scales, the latter being consistent with GS phenomenology.

Again, numerical simulations in the framework of EBM allowed us to interpret the change of local anisotropy as a result of the additional radial symmetry axis caused by expansion \citep{Verdini_Grappin_2015}.
These simulations reproduce the measured 3D anisotropy at all scales and further show that 
i) the large scale anisotropy is due to expansion which forces a variance anisotropy with radial symmetry axis \citep{Dong_al_2014}: the radial component of magnetic fluctuations decays faster than the transverse ones and fluctuations are confined in the plane transverse to the radial direction; 
ii) the scaling relations of $SF$ change when increments are taken along the radial direction (as in the above measurements exploiting single spacecraft data) or in directions transverse to the radial.
Both these numerical proofs were recently confirmed with a two-spacecraft analysis of solar wind data at scales of about $700-800~\mathrm{Mm}$ by \citet{Vech_Chen_2016}, who further found that: i) the solenoidality of the magnetic field also influences the ratio of transverse to radial magnetic fluctuations, ii) at small scales, $\sim 5-10~\mathrm{Mm}$, expansion effects are negligible, since the $SF$ is invariant under different sampling directions.

It is fundamental to notice that in the above EBM simulations (Fig.~5b in \citealt{Verdini_Grappin_2015}), when increments are taken transverse to the radial, the structure functions have three different power-law indices in the perpendicular, displacement, and parallel directions that define the local reference frame of the 3D local anisotropy.

This is in qualitative agreement with Boldyrev anisotropy \citep{B05,B06} in which the cascade timescale is slowed down by the tendency of velocity and magnetic fluctuations to become aligned at small scales (or equivalently of cross helicity to increase, see section~\ref{sec2} for its definition).
By requiring the alignment angle to be scale-dependent and using the critical balance, he obtained a non-axisymmetric anisotropy: $SF\sim \ell_\bot^{1/2}$ for increments perpendicular to the plane defined by the mean field and the fluctuation, $SF\sim {\cal L}_\bot^{2/3}$ for increments ${\cal L}_\bot$ in the fluctuation direction (displacement direction hereafter), and $SF\sim \ell_\|$ for the field-parallel direction.
According to this phenomenology, turbulent eddies become sheet-like structures (ribbons) at small scales, \chb{as suggested by intermittency analysis on MHD simulations \citep{Politano_al_1995} and solar wind data \citep{Carbone_al_1996}}.

Numerical simulations of MHD turbulence gave some evidence for the scale dependence of the alignment angle \citep{Beresnyak_Lazarian_2006,Mason_al_2006,Mason_al_2008,Beresnyak_Lazarian_2009,Perez_al_2012}, or for the three different power-law indices   \citep{Verdini_Grappin_2015,Mallet_al_2016}, but a direct confirmation from solar wind data is still missing (see \citealt{Podesta_al_2009,Wicks_al_2013} for a measure of the alignment angle).

To summarize, there are fundamental differences in the SF anisotropy when the sampling is in the radial direction or transverse to it, which is the consequence of the expansion-induced radial symmetry axis.  Such symmetry is absent at small enough scales ($10~\mathrm{Mm}$), but certainly present at large scales ($700~\mathrm{Mm}$), leaving at least one decade of inertial-range scales that are possibly subject to expansion.
In conclusion the only way to measure a dynamical regime that is controlled by the magnetic field axis is to minimize as much as possible the effects of expansion.
Thus, in this work we apply a selection criterium on solar wind data to obtain a measure of local anisotropy which is expected to be free of expansion effects. 
In particular we want to see if an observational proof of Boldyrev anisotropy can be found or if GS anisotropy persists once expansion effects are minimized.

The selection criterium is based on the above EBM simulations \citep{Verdini_Grappin_2015}. In particular, they showed that 
for measurements along the radial direction the large-scale local anisotropy $SF(\ell_\|)/SF({\cal L}_\bot)>1$ is controlled by the variance anisotropy $(b_{T}^2+b_N^2)/b_R^2>2 $ (the subscripts refer to the RTN coordinates, and $b$ is the root-mean-square amplitude of magnetic fluctuations, a ratio equal to $2$ corresponds to isotropy), both ratios growing with distance from the Sun.  
Interestingly, the variance anisotropy is not limited to large scales but it is about constant all the way down to dissipative scales, suggesting that expansion affects local anisotropy even in the inertial range of solar wind turbulence. 
We thus select intervals in which expansion is negligible by requiring a ratio of transverse to radial components smaller than two, and compute $SF$ to obtain the local anisotropy.

The plan of the paper is as follow. 
In section~\ref{sec2} we describe the data used (11 years of WIND data) and the selection method. 
The analysis method and the quantities that will be measured (structure functions and alignment angles) are given in section~\ref{sec3}.
We present the results on the 3D anisotropy along with the measurement of the alignment angle in section \ref{sec4}.
Section \ref{sec5} contains a summary and a discussion of the results.


\section{Dataset and Selection Criterion}\label{sec2}
\begin{figure}[t]
\begin{center}
\includegraphics [width=\linewidth]{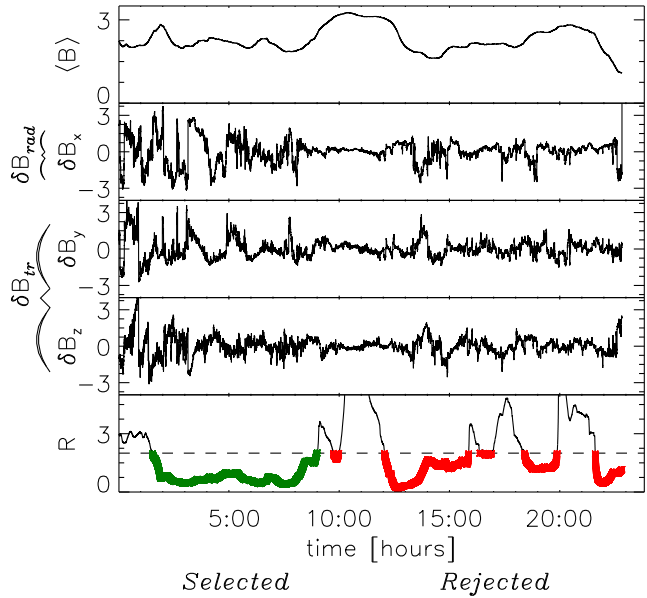}
\caption{Example of selected and rejected intervals in WIND data for the weak-expansion dataset according to eq.~\eqref{crit2}.  From top to bottom, mean magnetic field, magnetic fluctuations along the $X,~Y,~Z$ directions in GSE coordinate (all expressed in nT), and the ratio $R=\delta B_{rad}/\delta B_{tr}$ of radial to transverse rms amplitudes, which is used to select intervals with weak expansion. The selection criterium $R<2$ must be satisfied for 5 contiguous hours for an interval to be retained, as indicated by the green symbols in the bottom panel.}
\label{fig2}
\end{center}
\end{figure}
We analyze magnetic field data and plasma data at 1AU from instruments on WIND spacecraft in the period 2005-2015.  We use magnetic field data at $0.1s$ resolution from MFI instrument \citep{Lepping_al_1995} and onboard ion moments at $3s$ resolution from 3DP/PESA-L \citep{Lin_al_1995}.
To select intervals with weak or strong expansion we use the ratio 
\be
R=\frac{b_{tr}^2}{b_{rad}^2}=\left.\frac{b_Y^2 + b_Z^2}{b_X^2}\right|_{{\cal T}=2h}
\label{crit}
\ee
where $(X,Y,Z)$ are the GSE coordinates (with X aligned with the radial direction), the subscript $rad$ and $tr$ refer to the radial and transverse-to-the-radial components, and $b=\langle B-\langle B\rangle_{\cal T}\rangle_{\cal T}$ denotes the rms of fluctuations with respect to the mean field $\langle B\rangle_{\cal T}$, both computed at a timescale ${\cal T}$.
The mean and fluctuating magnetic fields are obtained by computing running averages on a window of duration ${\cal T}=2h$, which is about twice the convection time of the correlation length of turbulent fluctuations at 1AU \citep{Matthaeus_al_2005} and allows us to keep a trace of expansion effects.
We separate intervals with strong and weak expansion as follows:
\ba
2<R&<&10\;\; \mbox{strong-expansion dataset} \label{crit1}\\
0<R&<&2\;\;  \mbox{weak-expansion dataset} \label{crit2}
\ea
so that for strong expansion fluctuations are mainly transverse to the radial direction, while for weak expansion fluctuations are isotropic or mainly radial.

We require the criteria eqs.~\eqref{crit1}-\eqref{crit2} to be satisfied continuously for at least 5h, with a tolerance of 1 minute for possible data gaps or out-of-bounds R. An example of selected and rejected intervals for weak expansion is given in Fig.~\ref{fig2}, where the mean field, the fluctuations, and the ratio R are plotted as a function of time. 

In Table~\ref{table1} we provide some informations about the two datasets.
The strong-expansion dataset is about three times larger than the weak-expansion one, both in number of intervals and in total duration. The maximal duration of intervals is slightly smaller than 1 day for the strong-expansion dataset, a bit larger than half a day for the weak-expansion one.  On average all the analyzed intervals last about 6-7h. 
The two datasets have some common average properties, like the relative magnetic and density fluctuations ($0.7$ and $0.1$ respectively), or solar wind speed ($\approx 450~\mathrm{km/s}$). These are obtained as the mean of the mean in each interval. The distribution of the mean in each dataset is also similar (not shown). 

\begin{table*}[t]
\centering
\caption{Statistical properties of the two datasets with strong and weak expansion.}
\begin{tabular}{l|cccccc|ccccccc}
\hline
Dataset & 
$R$ & $N_{int}$ & $\Delta T|_{tot}$ & $\Delta T|_{max}$ & $\langle \Delta T\rangle$ & median $\Delta T$ & 
$\langle R\rangle$ & $\langle b_{rms}\rangle$ & $\langle b_{rms}/B_0\rangle$ & $\langle \rho_{rms}/\rho_0\rangle$ & $\langle V_{SW}\rangle$ & $\langle\sigma_c\rangle$ & $\langle\theta_{BR}\rangle$ \\
\hline\hline
Strong & $\in[2,10]$   & 1508 & 450d &  22h & 7.2h & 6.4h & 
6.3 & 3.0 nT& 0.72  & 0.11 &  482 km/s&  0.53 & $51^o$ \\
Weak   & $\in[0,2]$    &  564 & 157d &  16h & 6.7h & 6.1h & 
1.2 & 2.0 nT& 0.71  & 0.10 &  436 km/s&  0.13 & $74^o$ \\
\hline\hline
\end{tabular}
\label{table1}
\tablecomments{$R$ is the ratio of transverse to radial fluctuation energy used to separate the two datasets eq.~\eqref{crit}. $N_{int}$ is the number of intervals in each dataset, $\Delta T$ is the duration of a single interval. The total duration of the dataset is indicated in days, while the maximal, mean, and median durations of intervals are reported in hours.
In the last seven columns all quantities are the mean of the mean values computed in each interval \chb{(we first evaluate a quantity at 2h scale and then compute its mean value in a given interval)}. $b_{rms},~\rho_{rms}$ are the root mean square value of magnetic and density fluctuations in each interval, $B_0,~\rho_0$ are their mean value, $V_{SW}$ is the solar wind speed, $\sigma_c$ is the normalized cross helicity, and $\theta_{BR}$ is the angle of the mean field with respect to the radial direction.}
\end{table*}

\begin{figure}[t]
\begin{center}
\includegraphics [width=\linewidth]{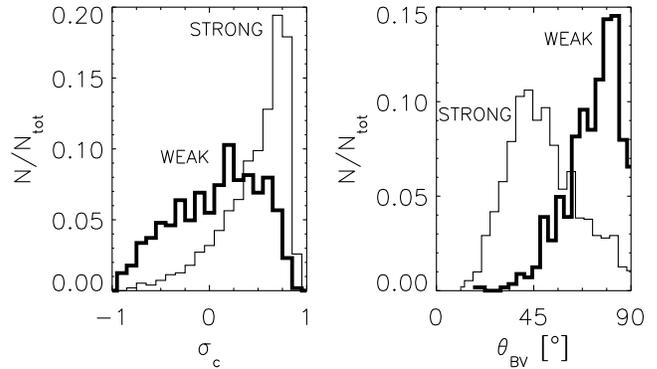}
\caption{Distribution of the cross helicity ($\sigma_c$, left panel) and of the angle between the mean magnetic field and the solar wind velocity ($\theta_{BV}$, right panel) averaged in each interval for the strong- and weak-expansion datasets (thin and thick lines, respectively).} 
\label{fig11}
\end{center}
\end{figure}

The only noticeable difference is in the value of the cross-helicity, $\sigma_c=-2\vect{u}\cdot\vect{b}/(\vect{u}^2+\vect{b}^2)$, and of the angle of the mean field with respect to the radial $\theta_{BR}$. 
In Fig.~\ref{fig11} we plot the distribution of the mean of $\sigma_c$ (left panel), and of $\theta_{BR}$ (right panel) calculated in each interval for both dataset (thin and thick lines for the strong- and weak-expansion datasets, respectively).
The high cross helicity is a prominent feature of the strong-expansion dataset 
while in the weak-expansion dataset the distribution is flatter and centred around zero (left panel)
On the other hand, the mean magnetic field is distributed around the Parker spiral in the strong-expansion dataset, while a transverse mean field is a characteristic of the weak-expansion dataset (right panel). 
We will come back on these different properties in the discussion.

\section{Data Analysis}\label{sec3}
\begin{figure}[t]
\begin{center}
\includegraphics [width=\linewidth]{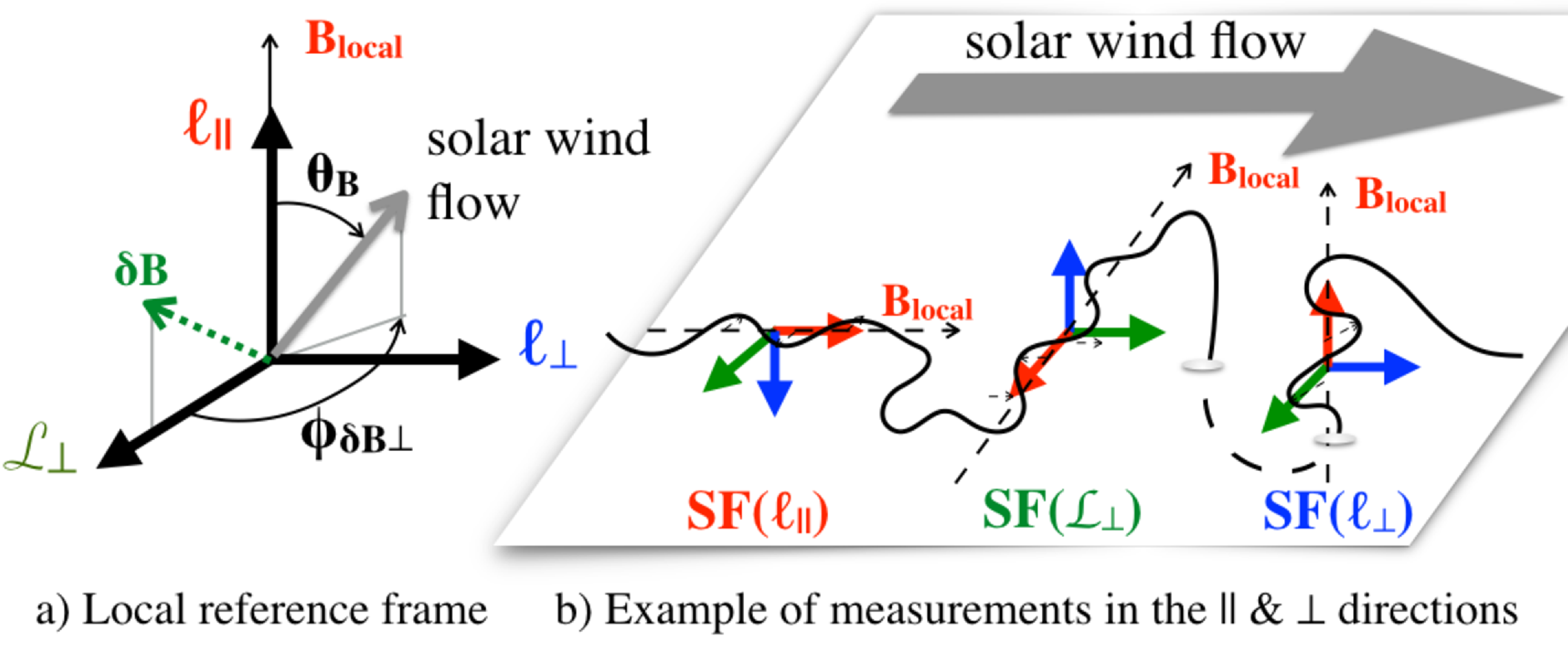}
\caption{a) Local reference frame used to compute structure functions \citep{Chen_al_2012}. b) An example of a field line in the solar wind flow that returns measurements along the three orthogonal directions defining the local reference frame: parallel (red), displacement (green), and perpendicular (blue) directions.}
\label{fig3}
\end{center}
\end{figure}

Following \citet{Chen_al_2012}, we compute the local structure function in each interval as follows.
For each pair of magnetic field $\vect{B}_1=\vect{B}(t)$, $\vect{B}_2=\vect{B}(t+\tau)$ separated by a time lag $\tau$, we compute the fluctuation, 
\be
\delta \vect{B}=\vect{B}_1-\vect{B}_2,
\label{deltaB}
\ee 
the local mean field, 
\be
\vect{B}_l=1/2(\vect{B}_1+\vect{B}_2),
\label{Blocal}
\ee 
and the local perpendicular displacement direction,
\be
\delta\vect{B}_\bot=\vect{B}_l\times[\delta\vect{B}\times\vect{B}_l].
\label{deltaBperp}
\ee
The latter two define the local reference frame axes $(\vect{e}_z,~\vect{e}_x)\equiv(\ell_\|,~{\cal L}_\bot)$, respectively, while the $\vect{e}_y\equiv\ell_\bot$ axis is orthogonal to both the fluctuation and the mean field. 
We use a spherical polar coordinate system (see Fig.~\ref{fig3}a) in which the radial vector $\vect{\ell}$ coincides with the solar wind flow direction, i.e. the sampling direction, and the polar $\theta_B$ and azimuthal $\phi_{\delta B\bot}$ angles measure deviations from the mean-field and the displacement directions, respectively.
We use two definitions of the solar wind speed and accordingly two different bin sizes when analyzing data at $0.1$ and $3s$ resolution, for reasons explained below.

When employing full resolution $0.1s$ MFI data, we use 66 logarithmically spaced increments to measure the power level in the range 
$10^{-3}~\mathrm{Mm^{-1}}<\kappa<10~\mathrm{Mm^{-1}}$, where $\kappa=1/\ell$ is the wavenumber obtained from the increment $\vect{\ell}=\tau \vect{V}_{SW}$. 
The sampling direction is given by the solar wind speed, $\vect{V}_{SW}$, which is the average of the first moment of the ion distribution computed in each interval, 
\be
\vect{V}_{SW}=\langle\vect{V}\rangle.
\label{vsw1}
\ee
For the polar and azimuthal angles we use $5^o$ bins to cover one quadrant only (any angle greater than $90^o$ is reflected below $90^o$).

The average angle between the solar wind speed and the radial is about $3^o$ with a maximal deviation of $12^o$ for some intervals. This is larger than our angular bin size and can affect the determination of the local mean field direction.

In fact, consider a rectangular eddy of aspect ratio $\ell_\bot/\ell_\|$. 
The angle $\Omega=\tan^{-1}(\ell_\bot/\ell_\|)$ is the angular measurement of its anisotropy.
A sampling with angular bin size  $\delta\theta_B<\Omega$ returns the parallel $SF$, while for larger $\delta\theta_B$ one starts to sample the perpendicular $SF$. 
If $\delta\theta_B$ is determined with an error larger than $\Omega$, the parallel $SF$ is contaminated with measurements of the perpendicular $SF$. 
This may happen because the sampling direction (the solar wind direction) is either not well determined or not well resolved, or because the mean field direction suffers from the same uncertainty. Given the relatively small scales considered and the definition of local mean field, eq.~\eqref{Blocal}, we expect the uncertainty in its direction to be small \citep{Gerick_al_2017} compared to that one of the solar wind direction.

To improve the accuracy in determining the solar wind direction, we adopt a ``local'' definition of the solar wind speed,
\be
\vect{V}_{SW}(\ell)=1/2(\vect{V}_1+\vect{V}_2),
\label{vsw2}
\ee
and we use a $1^o$ bin size for the polar and azimuthal angles. 
These choices allow us to determine the angle between the flow direction and the local magnetic field with better accuracy. 
Since ion moments have lower resolution we interpolate MFI data on $3s$ ion data before computing $SF$s and we use again 66 logarithmically spaced increments, but in the range 
$10^{-4}~\mathrm{Mm^{-1}}<\kappa<1~\mathrm{Mm^{-1}}$.

%
%

In order to obtain a structure function for a given dataset, we proceed as follows.
First we compute the structure function in each interval in the dataset, $SF_i$. 
Since the rms of fluctuations varies substantially from one interval to the other,  with $b_{rms}\in[0.3,18]~\mathrm{nT}$ in the weak-expansion dataset, and $b_{rms}\in[0.4,11]~\mathrm{nT}$ in the strong expansion dataset, we need to define a normalisation before averaging among intervals belonging to the same dataset. 
We thus select a scale $\ell^*$ and normalize the local $SF_i$ with the energy of fluctuations at that scale. To choose the normalization scale, we compute the global structure function in each interval, $S_i(\ell)$, i.e. the equivalent of a Fourier spectrum, and identify the range of scales in which all the $S_i$ have a power-law behavior. As far as $\ell^*$ belong to this common power-law range, the normalization does not affect the scaling relation that will be shown. We will use $\ell^*=10^2~\mathrm{Mm}$ and the normalisation energy is then $S_i(\ell^*)$ (we actually checked that results are unchanged when $\ell*$ varies by a factor 1/10).
Below, we give the definitions of $SF_i$ and $S_i$ and explain in detail the normalisation method.

The contribution to the power of fluctuations of each pair, $|\delta \vect{B}_j|^2$ is accumulated on the above-defined grids of increments and then the mean value in each bin is computed to obtain the structure function for a given selected interval $i$:
\be
SF_i(\ell,\theta_B,\phi_{\delta B\bot})=\frac{\sum_j\left|\delta \vect{B}_j(\ell,\theta_B,\phi_{\delta B\bot})\right|^2}{n_i(\ell,\theta_B,\phi_{\delta B\bot})},
\label{SFi}
\ee
where $n_i=\sum_j n_j$ is the number of measurements for each bin in the analyzed interval $i$.

The power of fluctuations at scale $\ell^*$ is simply the value of the global structure function $S_i(\ell^*)$, which is obtained by summing all the angular bins of the local structure function $SF_i$ and then by averaging
with weight $n_i/N_i$, where $N_i=\sum_{\theta_B,\phi_{\delta B\bot}}n_i$ is the total number of measurements falling in a given distance bin:
\be
S_i(\ell)=
\sum_{\theta_B,\phi_{\delta B\bot}}\left(SF_i(\ell,\theta_B,\phi_{\delta B\bot}) \frac {n_i(\ell,\theta_B,\phi_{\delta B\bot})}{N_i(\ell)}\right).
\label{SG}
\ee
We chose $\ell^*=10^2~\mathrm{Mm}$, which corresponds to a frequency range $3~10^{-3}~\mathrm{Hz}<f<8~10^{-3}~\mathrm{Hz}$ for an average solar wind speed $300~\mathrm{km/s}<V_{sw}<800~\mathrm{km/s}$, thus falling in the inertial range of solar wind turbulence \citep[e.g.][]{Podesta_al_2007,Chen_al_2013}.

Ultimately, the normalized local structure function is obtained as a weighted average of $SF_i/S_(\ell^*)$, with weight given by the count in each bin for a given interval, $n_i$, divided by the count in each bin for the whole dataset, $n=\sum_i n_i$: 
\be
SF(\ell,\theta_B,\phi_{\delta B\bot})=\sum_i \frac{SF_i(\ell,\theta_B,\phi_{\delta B\bot})}{S_i(\ell^*)}\times
\frac{n_i(\ell,\theta_B,\phi_{\delta B\bot})}{n(\ell,\theta_B,\phi_{\delta B\bot})}.
\label{SF}
\ee
In a similar way the normalized global structure function is 
\be
S(\ell)=\sum_i \frac{S_i(\ell)}{S_i(\ell^*)}\frac{N_i(\ell)}{N(\ell)},
\ee
where the weight is $N_i/N$ with $N=\sum_i N_i$ being the total dataset count in each distance bin.


We finally give the definitions of the alignment angle between perpendicular components of fluctuations 
which will be used to test Boldyrev phenomenology. Four definitions are given, which are the most used for in in-situ data and numerical data. 
The proper definition for the angle between velocity and magnetic fluctuation is  
\be
\sin\theta^{ub}_\bot\equiv\left\langle\frac{\delta\vect{B_\bot}\times\delta\vect{u_\bot}}{|\delta\vect{B_\bot}||\delta\vect{u_\bot}|}\right\rangle\label{th1},
\ee
where the angular brackets stand for a time average in the selected interval, the subscript $\bot$ indicates the component perpendicular to the local mean field, and $\delta u$ is defined as in eqs.~\eqref{deltaB},~\eqref{deltaBperp}. A similar definition can be given for the two Elsasser fields, 
\be
\sin\theta^{\,z}_\bot\equiv\left\langle\frac{\delta\vect{z^+_\bot}\times\delta\vect{z^-_\bot}}{|\delta\vect{z^+_\bot}||\delta\vect{z^-_\bot}|}\right\rangle,
\label{th2}
\ee
with $\delta\vect{z}^\pm=\delta\vect{u}\pm \delta\vect{B}/\sqrt{4\pi\rho_0}$.
This angle is related to the residual energy (the excess of magnetic to kinetic energy) and not to cross-helicity. It can be thought as a correction to the order of magnitude estimate of nonlinear terms based on reduced MHD\footnote{In incompressible MHD the nonlinear terms are proportional to $\vect{z}^\pm\cdot\bnabla\vect{z}^\mp$. In reduced MHD, both wavevectors and fluctuations lie in the perpendicular plane, so to the lowest order nonlinearities are $\sim k_\bot z_\bot^+z_\bot^-\sin\theta^{\,z}_\bot$, where the angle between $z^\pm$ accounts for the weakening introduced by this geometrical constraint.}. 
It is worth noticing that Boldyrev theory is based on the increase of cross helicity at smaller scale, which is related the angle between velocity and magnetic fluctuations and not to that one between Elsasser fields.

Measurements of the above angles in numerical simulations \citep{Beresnyak_Lazarian_2009} displayed virtually no scaling with increments.
A different definition obtains by averaging separately the numerator and the denominator, in what is termed polarization intermittency \citep{Beresnyak_Lazarian_2006}:
\be
\sin\tilde{\theta}^{ub}_\bot\equiv\frac{\left\langle\delta\vect{B_\bot}\times\delta\vect{u_\bot}\right\rangle}{\left\langle|\delta\vect{B_\bot}||\delta\vect{u_\bot}|\right\rangle},
\label{th3}
\ee
and analogously,
\be
\sin\tilde{\theta}^{\,z}_\bot\equiv\frac{\left\langle\delta\vect{z^+_\bot}\times\delta\vect{z^-_\bot}\right\rangle}{\left\langle|\delta\vect{z^+_\bot}||\delta\vect{z^-_\bot}|\right\rangle}.
 \label{th4}
\ee
These latter definitions, eqs~\eqref{th3},~\eqref{th4}, returned scaling relations compatible with Boldyrev phenomenology,
\be
\sin\theta_\bot\sim\ell_\bot^{1/4}\label{eq:angle},
\ee
in numerical simulations \citep{Mason_al_2006, Beresnyak_Lazarian_2006, Mason_al_2008,Beresnyak_Lazarian_2009,Perez_al_2012} and in some solar wind intervals \citep{Podesta_al_2009}. 
The definition involving Els\"ass\"er fields
was also analyzed in numerical simulations of turbulence in the acceleration region of the solar wind \citep{Perez_Chandran_2013} and was used in \citet{Mallet_al_2016} to test the critical balance applied to Boldyrev phenomenology. 

Since both velocity and magnetic fluctuations are needed to measure the above angles, we interpolate $0.1s$ MFI data on $3s$ ion moments. 
The density entering the definition of Elsasser variables is the average density in each interval, and we use eq.~\eqref{vsw2} to determine the sampling direction. 
For each interval we obtain the angles $\theta_{\bot,i}$ defined in eqs.~\eqref{th1}-\eqref{th4} as a function of the perpendicular scale $\ell_\bot$ (where $\ell_\bot$ is perpendicular to $\delta\vect{B}_\bot$ and to $\delta \vect{z}^+_\bot$ for the angles in eqs.~\eqref{th1},~\eqref{th3} and in eqs.~\eqref{th2},~\eqref{th4} respectively, see Fig.~\ref{fig3}). We then compute the angles for the dataset by averaging over intervals $\theta_\bot=\langle\theta_{\bot,i}\rangle$.

\begin{figure}[t]
\begin{center}
\includegraphics [width=\linewidth]{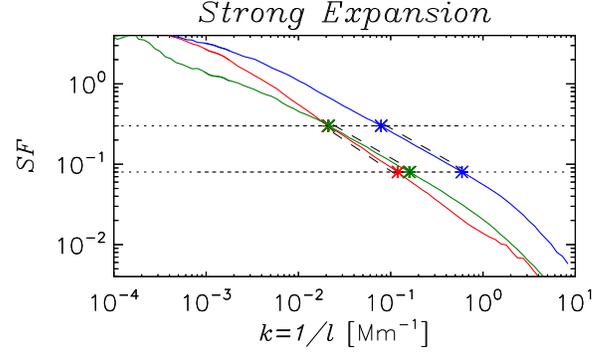}
\caption{Local structure functions for the strong-expansion dataset. Blue, green, and red colors indicate the perpendicular, displacement, and parallel directions respectively. The dashed horizontal lines bracket the energy intervals in which power-law indices are measured, the fitted power laws $SF~\sim \ell^\alpha$ are also drown with thin black lines, with exponents $\alpha=0.65,~0.65,~0.77$ for the perpendicular, displacement, and parallel directions respectively.}
\label{fig5}
\end{center}
\end{figure}

\section{Results}\label{sec4}
We first analyze \textit{local} $SF$ of the strong-expansion dataset. 
We use full resolution $0.1s$ MFI data with the definition of the solar wind speed ~\eqref{vsw1}, and focus on the scaling relations along the perpendicular, displacement, and the parallel $SF$s, which correspond to the following angular bins:
\ba
SF(\ell_\bot)&\rightarrow&(85^o<\theta_B<90^o,85^o<\phi_{\delta B\bot}<90^o),
\label{per}\\
SF({\cal L}_\bot)&\rightarrow&(85^o<\theta_B<90^o,0^o<\phi_{\delta B\bot}<5^o),
\label{flu}\\
SF(\ell_\|)&\rightarrow&(0^o<\theta_B<5^o,0^o<\phi_{\delta B\bot}<90^o).
\label{par}
\ea
These measurements reflect particular configurations of the field-line and sampling directions as represented in Fig.~\ref{fig3}b.
Note that the definition of $SF(\ell_\|)$ is slightly different from the one used in \citet{Chen_al_2012}: we average over the angle $\phi_{\delta B \bot}$ instead of using the bin $0^o<\phi_{\delta B\bot}<5^o$.

The analysis of the weak-expansion dataset follows. 
In this section we also make use of ion data at $3s$ resolution to determine the solar wind speed with more accuracy (see the definition of solar wind velocity, eq.~\eqref{vsw2}) and to measure the alignment angles, eqs.\eqref{th1}-\eqref{th4}. When ion data are involved, we will interpolate the $0.1s$ MFI data on their timestamps.

We then show \textit{global} structure functions, which, contrary to local structure functions, have very similar properties in the two datasets.

Finally, we describe in more detail the angular distribution of the anisotropy in the weak-expansion dataset, that allows us to reconcile the properties of local and global structure functions.

\subsection{Strong-expansion dataset, local structure functions}

We plot the local structure function for the strong-expansion dataset in the three orthogonal perpendicular (blue), displacement (green), and parallel (red) directions in Fig.~\ref{fig5}.

An eddy can be identified with a given energy level of $SF$ and the anisotropy can be quantified by the measure of its characteristic scales in the above three orthogonal directions. For example, choosing the level corresponding to the upper dashed line in Fig.~\ref{fig5}, the eddy has a perpendicular scale much smaller than the parallel and displacement scales (the symbols on the same dashed line). 
In the following, to describe anisotropy we will refer to large and small energies to describe large- and small-scale anisotropy. When a scale is explicitly mentioned it indicates the smallest of the three scales defining the eddy anisotropy.

In Fig.~\ref{fig5}, we recover the anisotropy already observed by \citet{Chen_al_2012} in the fast solar wind with Ulysses data.
i) As a general feature, $SF(\ell_\bot)$ and $SF({\cal L}_\bot)$ have the same scaling, their ratio is about 2.8 at all scales.
ii) At large energy (above the upper dashed horizontal line) the parallel $SF$ is as energetic as the perpendicular one. This corresponds to eddies elongated in the displacement direction. 
iii) At smaller energy (within the dotted horizontal lines), the two perpendicular $SF$s dominate the parallel one.
As a result, eddies become more and more elongated in the parallel direction, with a constant aspect ratio in the perpendicular plane (tubes).

At scales larger than $1500~\mathrm{km}$ (i.e. in the energy range $0.08<SF<0.3$) the power-law indices $SF\sim\ell^{\alpha}$  are $\alpha=0.65,~0.66,~0.77$ for the perpendicular, displacement, and parallel directions. The two perpendicular $SF$ are close to each other and consistent with the 2/3 value also found in Ulysses data.
Note, however, that because of our definition, eq.\eqref{par}, the parallel $SF$ has a flatter power-law index compared to the value found with Ulysses data, $0.86$ \citep{Chen_al_2012}, both being flatter than the critical balance prediction (slope $1$).
\begin{figure}[t]
\begin{center}
\includegraphics [width=\linewidth]{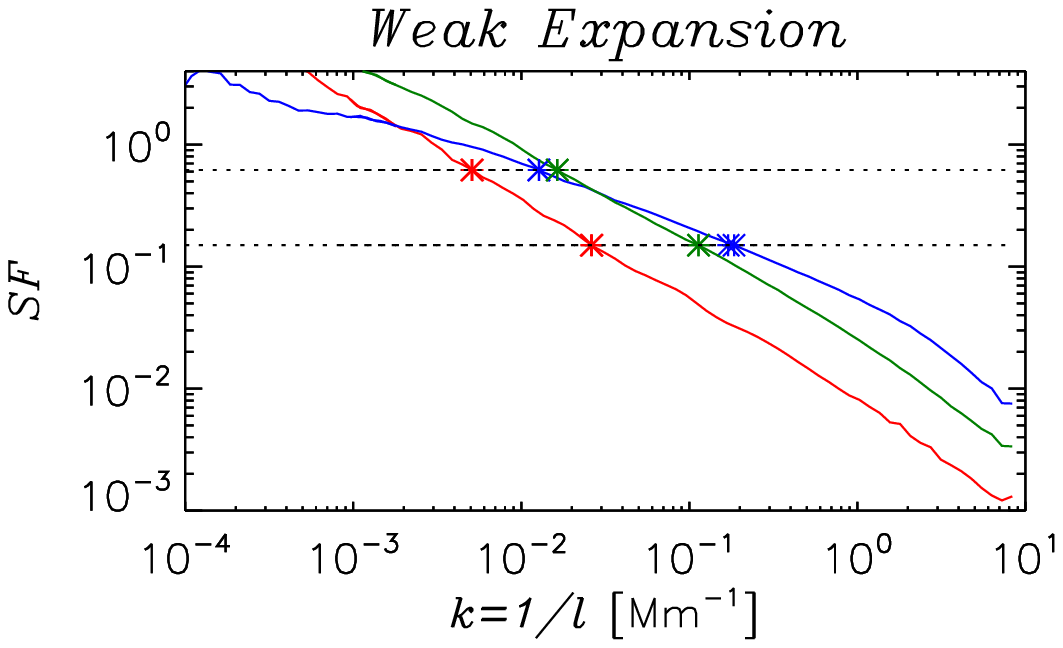}\\
\includegraphics [width=\linewidth]{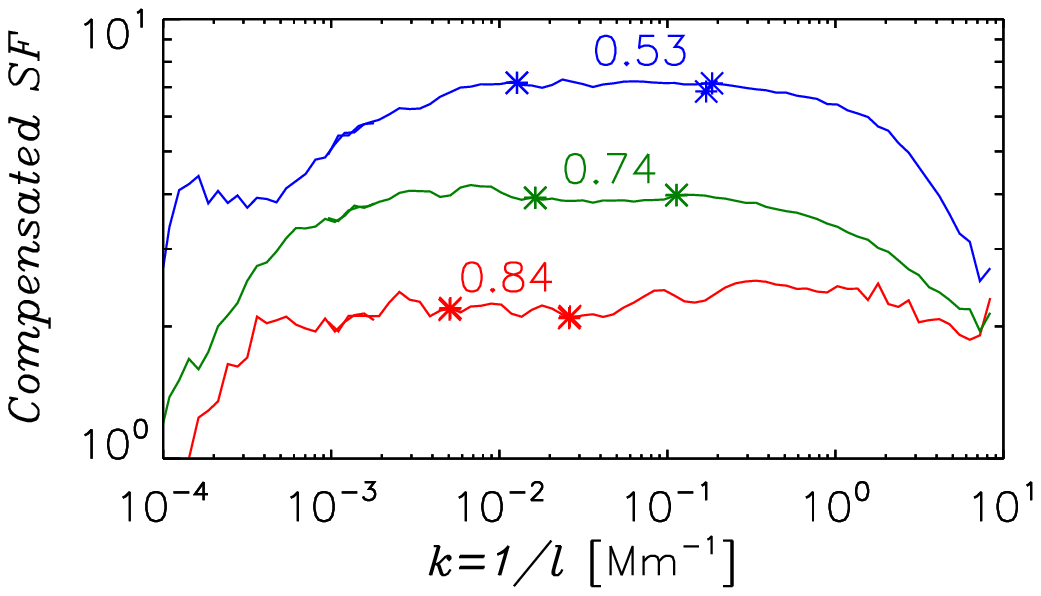}
\caption{Top panel. Local structure functions for the weak-expansion dataset. Blue, green, and red colors indicate the perpendicular, displacement, and parallel directions respectively. The dashed horizontal lines bracket the energy intervals in which power-law indices are measured. 
In the bottom panel the $SF$ are arbitrarily shifted in the vertical direction and compensated by $k^{\alpha}$, where $\alpha=0.52,~0.74,~0.84$ is obtained by fitting $SF\sim\ell^\alpha$ in the range delimited by symbols.}
\label{fig6}
\end{center}
\end{figure}

\begin{figure}[t]
\begin{center}
\includegraphics [width=\linewidth]{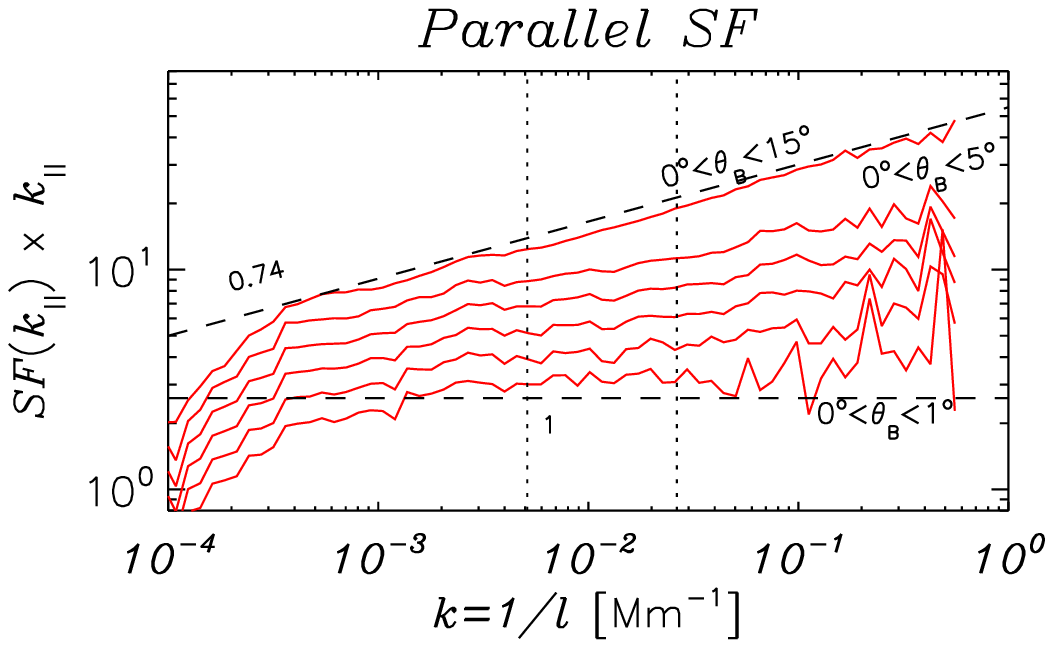}\\
\includegraphics [width=\linewidth]{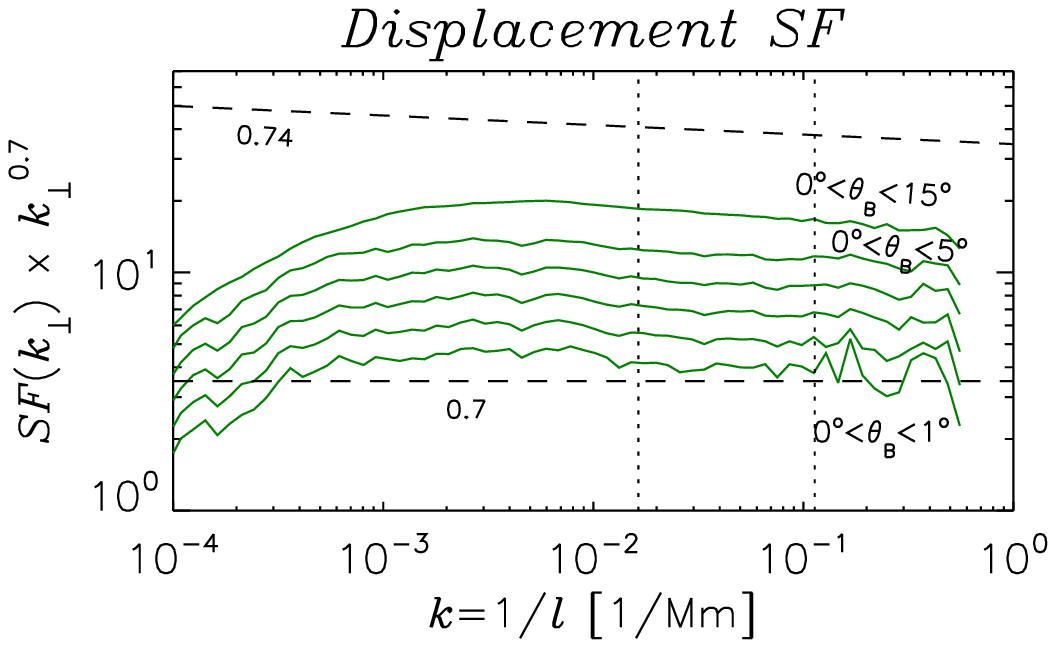}
\caption{Weak expansion dataset 2, structure functions obtained with bins size $\delta\theta_B=1^o$ with $0.1s$ MFI data interpolated on $3s$ ion moments, and using definition eq.~\eqref{vsw2} for the solar wind speed. 
Top panel, $SF(\ell_\|)$ compensated by $k$, bottom panel $SF({\cal L}_\bot)$ compensated by $k^{2/3}$.
In both panels the $SF$ are multiplied by an arbitrary factor so that the bin size increases by $1^o$ from bottom to top (except for the last line at $15^o$ bin). 
The dotted vertical lines indicate the range for fitting the power law index in Fig.~\ref{fig6}.
The dashed lines are references for the scaling $SF\sim\ell^\alpha$ with $\alpha$ indicated in the figures.}
\label{fig7}
\end{center}
\end{figure}

At scales smaller that $500~\mathrm{km}$ the three $SF$s steepen, which is suggestive of the high-frequency spectral break often observed in the magnetic field spectrum \citep[e.g.][]{Alexandrova_al_2012}. This is expected to occur \chb{at scales a factor two larger than the proton inertial scale ($\lambda_p$) and the proton Larmor radius ($\rho_p$) for plasma $\beta$ around one \citep{Bruno_Trenchi_2014} or at the largest of the two scales for low and high plasma $\beta$  \citep{Chen_al_2014,Franci_al_2016}}. In our dataset, these proton scales vary from one interval to the other, with mean and standard deviation $\lambda_p=130\pm90~\mathrm{km}$ and $\rho_p\approx90\pm40~\mathrm{km}$, both corresponding to the smallest resolved scale in Fig.~\ref{fig5} (similar values are found for the weak-expansion dataset). 
In our analysis that considers scales well above the proton scales, the steepening of $SF$ at $\ell\approx500~\mathrm{km}$ ($k\approx 2~\mathrm{Mm^{-1}}$) is likely due to filtering that have been applied to calibrate the data (a steepening at a scale of about 5 times the smallest resolvable scale is also found when using data calibrated at $3s$ resolution from CDAWeb).

\subsection{Weak-expansion dataset, local structure functions and alignment angle}

The local $SF$ for the weak-expansion dataset are shown in Fig.~\ref{fig6}.
The anisotropy is different from the previous case. 
i) At large energy, $SF({\cal L}_\bot)>SF(\ell_\bot)$. 
ii) At intermediate energies (between the dashed horizontal lines), the $SF$s have different power-law indices.
iii) At small energy (below the lower dashed horizontal line), the displacement and parallel $SF$ have almost the same scaling.

A fitting procedure in the energy range $0.15<SF<0.62$, corresponding to scales larger than $5500~\mathrm{km}$, returns the indices $\alpha=0.53,~0.74,~0.84$ for the perpendicular, displacement, and parallel $SF$ respectively, as can be seen in the bottom panel of Fig.\ref{fig6}, where $SF$ are compensated by $k^{\alpha}$ and arbitrarily shifted in the vertical direction. 
Note that the index of the perpendicular $SF(\ell_\bot)$ is close to Boldyrev phenomenology (1/2), in the displacement direction it is closer to $2/3$ than $1$, and it is significantly steeper in the parallel direction (although again smaller than 1).

The value of the power-law index in the parallel $SF$ is sensitive to determination of the sampling direction, which may be subject to error due to the variation of solar wind speed within each interval and to the chosen value of the angular bin size.
In the weak-expansion dataset, the angle quantifying the eddy anisotropy between the perpendicular and parallel directions is 
$\Omega=\tan^{-1}(k_\|/k_\bot)=\tan^{-1}(0.005/0.012)\sim8^o$ 
at the lowest horizontal dashed line in Fig.~\ref{fig6} (top panel, $SF=0.15$).
This value is close to our angular bin size ($\delta\theta_B=5^0$), and since the average deviation of the solar wind direction from the radial one is $3^o$, it can lead to a wrong determination of the field-parallel direction and the corresponding scaling law.

To test if this uncertainty affects the scaling of $SF$ we compute the solar wind speed locally, eq.~\eqref{vsw2}, and use a $1^o$ bin size to determine more precisely the sampling direction. $SF$s are now computed with MFI data interpolated onto $3s$ ion moments data,  so they will extend to lower $k$.
In Fig.~\ref{fig7} we show the compensated parallel and displacement structure functions, $SF(\ell_\|)\times k$ in the top panel and $SF({\cal L}_\bot)\times k^{0.7}$ in the bottom panel, obtained with increasing angular bin size. Since the energy of $SF$ does not vary with the bin size, for better readability we shifted the curve by an arbitrary factor in the vertical direction.
At the smaller bin size the power-law indices are consistent with the values expected for Boldyrev phenomenology ( $1$ and $2/3$ for the parallel and displacement $SF$, respectively).
This corresponds to eddies being more and more elongated in the parallel direction and with an increasing aspect-ratio in the perpendicular plane (ribbons)
\footnote{We do not observe changes in the scaling exponents of $SF(\ell_\bot)$ in the weak expansion dataset, and in all $SF$s of the strong-expansion dataset (not shown).}.

Note that increasing the bin size has the effect of flattening the $SF$s, with the power-law indices becoming equal to 0.74.
This may explain the findings of \citet{Wang_al_2016} who studied $SF$ for intervals with weak fluctuations, $b_{rms}/B_0<1/10$. 
In their analysis they assumed a radial solar wind speed and obtained a power-law index of $2/3$ in both the (axisymmetric) perpendicular and parallel directions. For such small fluctuations the eddy aspect ratio could be large, and the angle $\Omega$ measuring the eddy anisotropy could be smaller than the actual deviation of the solar wind speed from the radial direction.

\begin{figure}[t]
\begin{center}
\includegraphics [width=\linewidth]{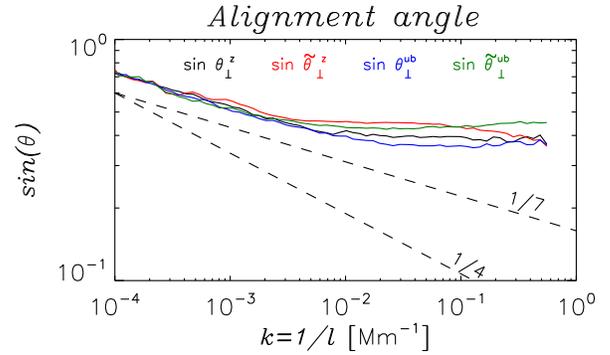}
\caption{Measeurement of the alignment angles eqs.~\eqref{th1}-\eqref{th4} in the weak-expansion dataset. The dashed lines are references for the scaling $\theta_\bot\sim\ell^\alpha$ with $\alpha$ indicated in the figure.}
\label{fig8}
\end{center}
\end{figure}
In order to test Boldyrev phenomenology, the scaling of $SF$s in the three orthogonal directions is complemented by the measurement of the alignment angle between the perpendicular components of fluctuations (we recall that only the angle between magnetic and velocity fluctuations, eq.~\eqref{th1} or eq.~\eqref{th3}, is involved in Boldyrev theory).
In Fig.~\ref{fig8} we plot the measurements of angles eqs.~\eqref{th1}-\eqref{th4} as a function of the perpendicular wavenumber, $k=1/\ell_\bot$, obtained with $0.1s$ MFI data interpolated on $3s$ ion moments and with $5^o$ angular bin size (the result is unchanged for $1^o$ bin size).
The four angles have similar scaling with $\ell_\bot$, at variance with numerical results \citep{Beresnyak_Lazarian_2009}, and most importantly they are not consistent with Boldyrev phenomenology.
First, the power law exponent ($\sim1/7$) is much smaller than what expected ($1/4$), and it is found at large scales $10^{-4}~\mathrm{Mm^{-1}}\lesssim k \lesssim 4~10^{-3}~\mathrm{Mm^{-1}}$. 
Second, the alignements angle is scale independent for $k\gtrsim 10^{-2}~\mathrm{Mm^{-1}}$, which is the range where$SF\sim\ell_\bot^{0.53}$ and $SF\sim{\cal L}_\bot^{0.74}$ (see Fig.~\ref{fig6})

\subsection{Global structure functions}
\begin{figure}[t]
\begin{center}
\includegraphics [width=\linewidth]{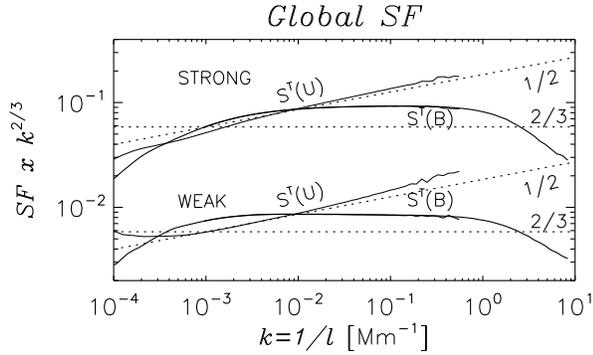}
\caption{Global structure functions compensated for $\ell^{2/3}$ for the two datasets. Lower lines refers to the strong-expansion dataset, upper lines to the weak-expansion dataset. Structure functions for magnetic and velocity fluctuations are in thick and thin lines, respectively. For the magnetic fluctuations we show both the $SF$ obtained with $0.1s$ MFI data and with MFI data interpolated on ion moments at $3s$ resolution. The dotted lines are references for $SF\sim\ell^\alpha$ with $\alpha=2/3$ and $1/2$ as indicated in the figure.}
\label{fig4}
\end{center}
\end{figure}
\begin{figure*}[t]
\begin{center}
\includegraphics [width=0.49\linewidth]{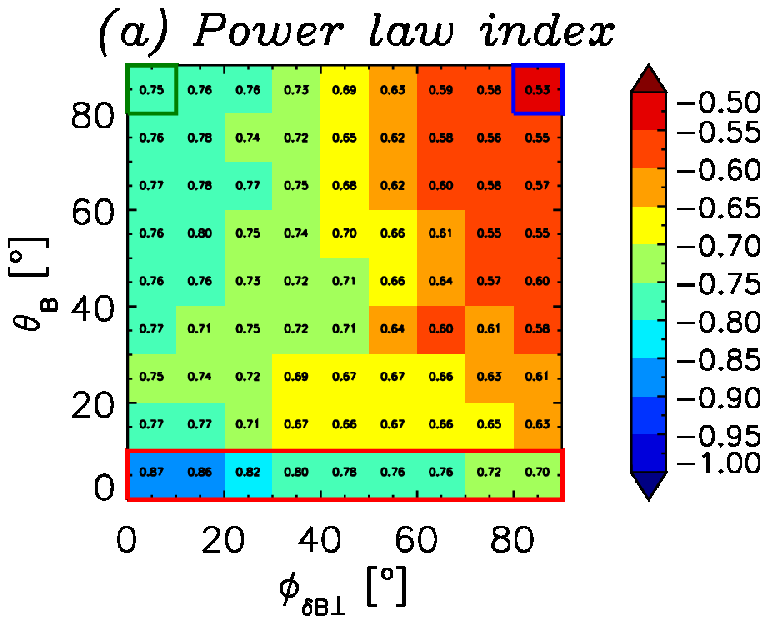}
\includegraphics [width=0.49\linewidth]{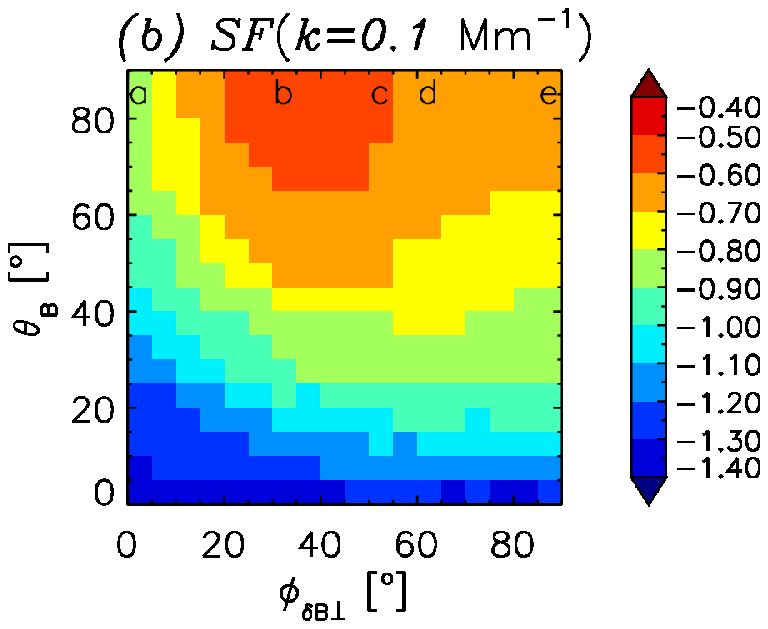}
\caption{Weak expansion dataset. 
Panel a): power law index $SF\sim\ell^{\alpha}$ measured in the energy interval $0.15<SF<0.62$ in all angular bins (reduced at $10^o$ for better readability). 
The red rectangle, the blue square, and the green square denote the directions $\ell_\|,{\cal L}_\bot,~\ell_\bot$, respectively. The red square in the bottom mark the parallel direction according to \citet{Chen_al_2012}.
Panel b): logarithmic energy levels in all angular bins at scale $k=0.1~\mathrm{Mm^{-1}}$. 
The letters in the top row mark the angular bins used to plot the $SF$ in Fig.~\ref{fig10a}.}
\label{fig10}
\end{center}
\end{figure*}

According to GS or Boldyrev theories, global structure functions should have the same scaling of the dominant direction in local structure functions. 
We thus expect to find the index 2/3 for strong expansion, and 1/2 for weak expansion (the blue lines in Figs.~\ref{fig5}-\ref{fig6}). 

The global structure functions for the magnetic and velocity fluctuations are plotted in Fig.~\ref{fig4}, compensated by $k^{2/3}$.
Surprisingly, both datasets have the same power-law index of 2/3 for the magnetic structure function (thick lines).
Note that the power-law behaviour extends for more than one decade in the magnetic $SF$. It starts at slightly larger scales in the weak-expansion dataset, possibly reflecting the weaker effect of expansion, while it ends at about the same scale. 
In both cases, the chosen normalisation scale, $\ell^*=100~\mathrm{Mm}$, is in the power-law range of the global $SF$. 

The power-law index of the velocity $SF$ (thin lines) is close to 1/2 (although a bit flatter for the weak-expansion dataset), and the extent of the region with power-law behaviour is about the same one of magnetic $SF$.

To understand why the weak-expansion dataset has a $2/3$ index for the magnetic structure function, we examine the angular distribution of anisotropy in the next section.

\begin{figure}[t]
\begin{center}
\includegraphics [width=\linewidth]{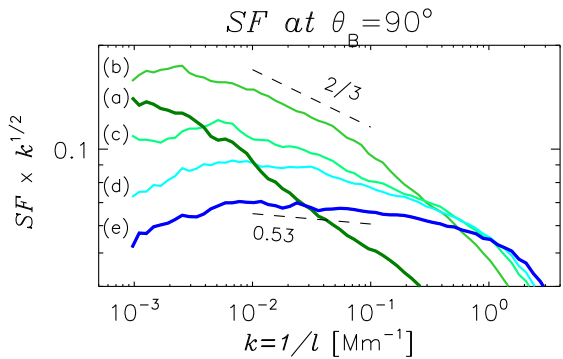}
\caption{Weak expansion dataset. 
Compensated $SF$ in the local perpendicular plane ($\theta_B=90^o$) for the azimuthal angles ($\phi_{\delta B\bot}$) corresponding to the letters in the top row of Fig.~\ref{fig10}b. Letters form (a) to (e) span increasing azimuthal angles, passing from the displacement (green) to the perpendicular (blue) directions.}
\label{fig10a}
\end{center}
\end{figure}

\subsection{Weak-expansion dataset: angular distribution of anisotropy}

In this section we use $SF$ computed with $0.1s$ MFI data, $5^o$ angular bin size, and the definition of the solar wind speed given in eq.\eqref{vsw1} (i.e. the same data used in Fig.~\ref{fig6}). 

In Fig.~\ref{fig10} we show the distribution in the plane ($\phi_{\delta B\bot},~\theta_B$) of the power-law index in the energy range $0.15<SF<0.62$ (left panel), and of the $SF$ energy at scale $k=0.1~\mathrm{Mm^{-1}}$ (right panel).
For better readability, in the left panel we have down-sampled the angular resolution to $10^o$ before the fitting procedure. 
In both figures the top right and top left corners correspond to the perpendicular and displacement directions, respectively, while the bottom line is the parallel direction. These directions are indicated with the blue (perpendicular), green (displacement), and red (parallel) boxes in the left panel.

In panel (a), the power-law index of $SF\sim\ell^\alpha$ has a monotonic increase from right to left (and to the bottom). When passing from the perpendicular direction to the displacement direction or to the parallel direction, the $SF$ steepens.

On the contrary the energy distribution in panel (b) is not monotonic when passing from the perpendicular to the displacement direction. 
The most energetic part of the $SF$ is perpendicular to the mean field (upper rows in the figure) at angles that are about $30^o$ away from the displacement direction.

This is better seen in Fig.~\ref{fig10a}, where we plot the $SF$ compensated by $k^{1/2}$ for some selected azimuthal angles ($\phi_{\delta B\bot}$) in the plane perpendicular to the mean field ($\theta_B=90^o$), corresponding to the postion of letters in the upper row of Fig.~\ref{fig10}b.
It is evident that the most energetic part of the $SF$ is in the range $20^o<\phi_{\delta B\bot}<40^o$, with steep power-law index $0.61<\alpha<0.75$, which explains why the global structure function has a power-law index of $2/3$ despite the value $\sim1/2$ found in the perpendicular direction.

\begin{figure*}[t]
\begin{center}
\includegraphics [width=0.35\linewidth]{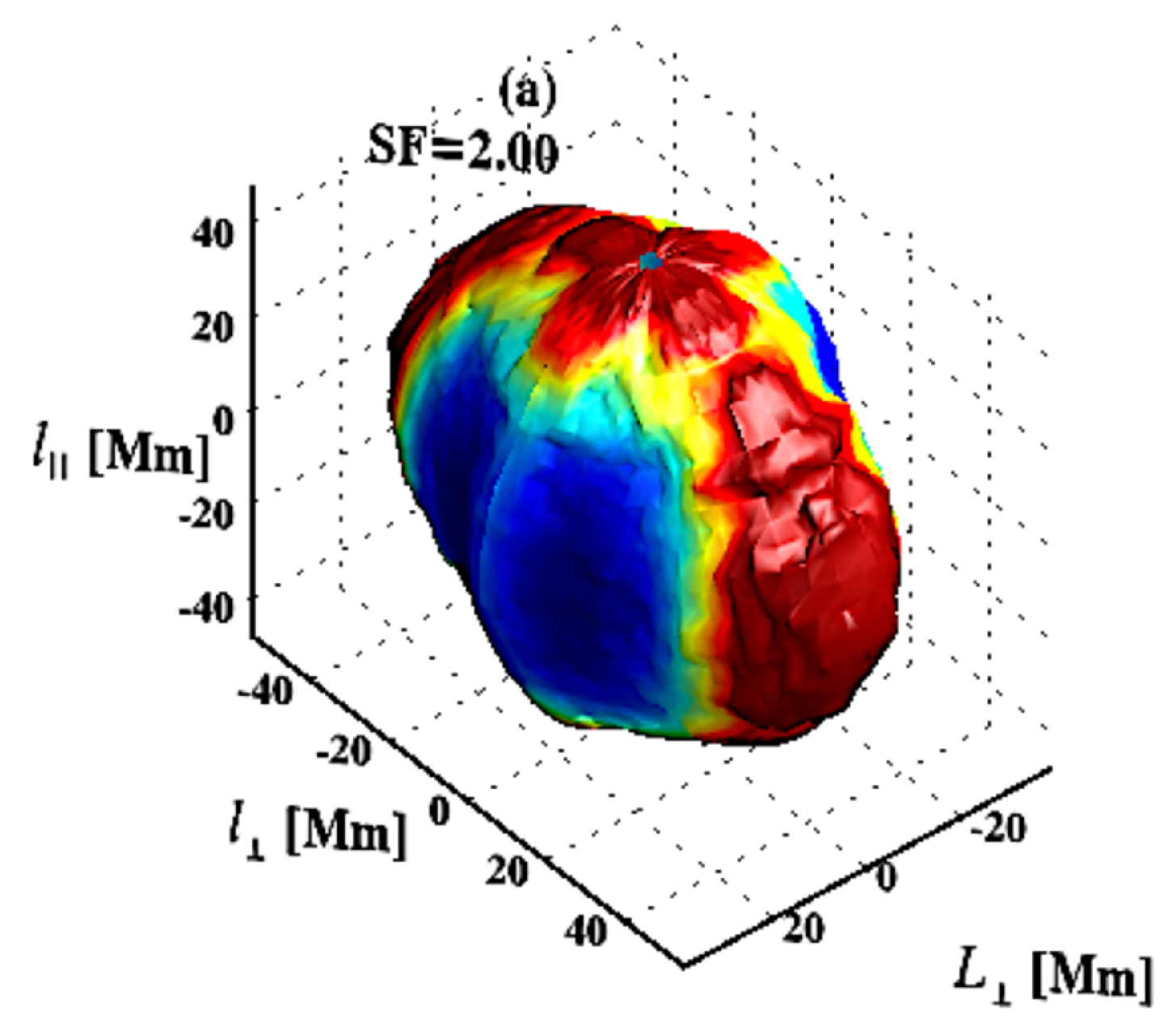}
\includegraphics [width=0.33\linewidth]{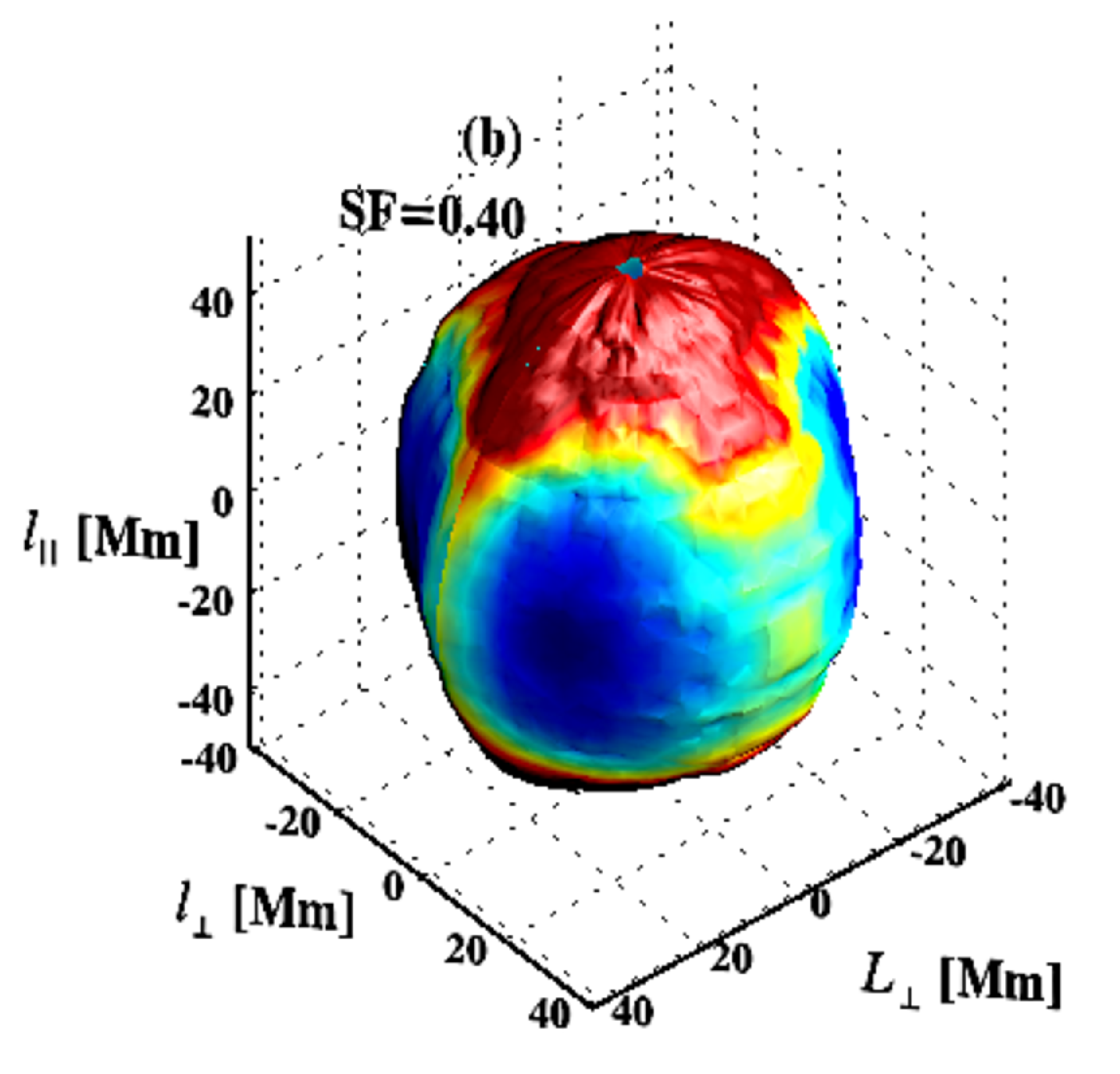}
\includegraphics [width=0.28\linewidth]{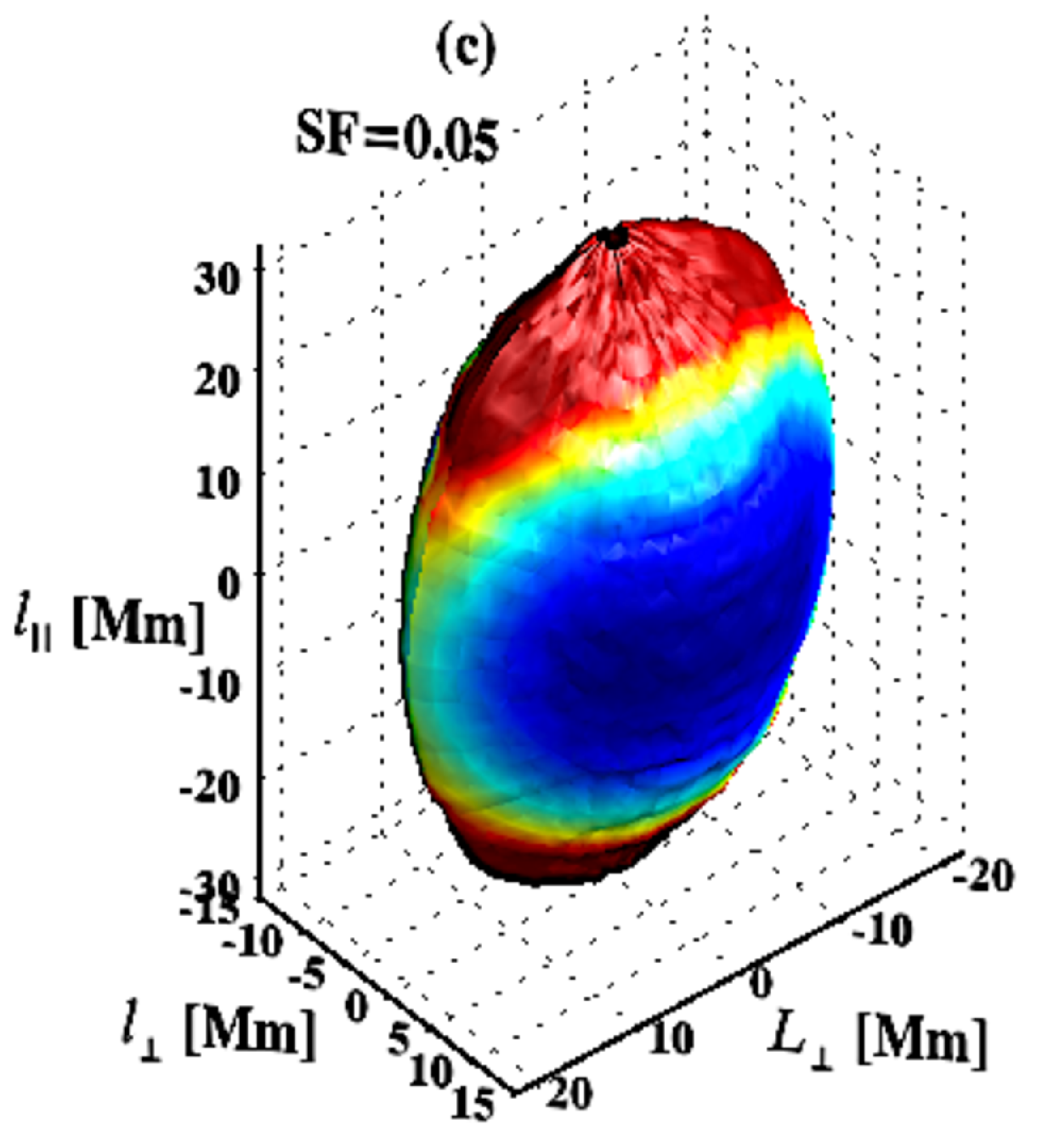}
\caption{Eddy shape in the weak-expansion dataset.
Isosurfaces of $SF$ for decreasing energy levels, roughly corresponding to decreasing scales.
The largest, intermediate, and smallest energies in panels a,b, and c correspond to levels above, in the middle, and below the power-law energy range delimited by the two dashed horizontal lines in the top panel of Fig.~\ref{fig6}. 
The colors are redundant: they indicate distances from the origin to help in visualizing the anisotropy.}
\label{fig12}
\end{center}
\end{figure*}

We finally give a 3D representation of the eddy anisotropy with scale for comparison with previous works \citep{Chen_al_2012, Verdini_Grappin_2015,Mallet_al_2016}.
In Fig.~\ref{fig11} we show isovolumes at constant energy for three values, $SF=2,~0.3,$ and $0.05$, corresponding roughly to decreasing scales. 

The largest energy (panel a) captures scales above the power-law region in Fig.~\ref{fig6}. At these scales, the eddy is elongated in the perpendicular direction, an anisotropy that differs from Ulysses observations \citep{Chen_al_2012} and numerical simulations with expansion \citep{Verdini_Grappin_2015} in which the eddy was elongated in the displacement direction.

The intermediate energy (panel b) is inside the energy interval in which power-law indices are measured (the dashed horizontal lines in the top panel of Fig.~\ref{fig6}). 
Here the eddy becomes elongated in the parallel direction and it is about axisymmetric with roughly equal dimensions in the perpendicular and displacement direction.

At energy below the power-law range interval (panel c), the eddy is 3D anisotropic, with the smallest dimension in the perpendicular direction, followed by the displacement direction, while the largerst dimension is parallel to the local mean field. Note that this shape is qualitatively similar to that one at small scales in the strong-expansion dataset.

The angular anisotropy of the strong-expansion dataset is almost identical to that one found by \citet{Chen_al_2012} and is not shown here: 
the spectral index is the same for the perpendicular and displacement directions, and it decreases toward the parallel direction;
the power decreases monotonically when moving from the perpendicular to the displacement and parallel directions; 
the eddy is elongated in the displacement direction at large scales and in the parallel direction at small scales. 
Note that with this monotonic distribution, the global structure function reflects the properties of the local structure function in the perpendicular direction, at variance with the weak-expansion dataset.

\section{Summary and Discussion}\label{sec5}
In this work we applied a selection on solar wind data that distinguishes intervals in which expansion effects are strong or weak.
We measure the anisotropy with respect to the local mean magnetic field in both samples with the aim of characterizing the anisotropy of strong MHD turbulence. This is expected to emerge in intervals with weak expansion in which the mean field is the only symmetry axis, while in intervals with strong expansion, both the radial direction and the magnetic field direction contribute to the symmetry properties of turbulence.
 
The selection criterium is based on numerical findings \citep{Dong_al_2014,Verdini_Grappin_2015,Verdini_Grappin_2016} that showed how the large-scale local anisotropy is controlled by the variance anisotropy of magnetic fluctuations, which tend to be confined in the plane perpendicular to the radial as a result of the solar wind expansion. 

We use MFI data and ion moments data from WIND spacecraft at 1~AU to compute local structure functions in two datasets that differ in their variance anisotropy $R=b_{tr}/b_{rad}$, where the rms fluctuations, $b$, are computed at $2h$ scale (here the subscripts $rad$ and $tr$ indicate the radial and transverse-to-the-radial directions).

The strong-expansion dataset has $2<R<10$. We recover quantitatively the anisotropy already obtained with Ulysses data \citep{Chen_al_2012}, with spectral indices $2/3,~2/3,~0.77$ for the perpendicular, displacement, and parallel direction. Note that the parallel direction is less steep than in Ulysses data because of our definition, eq.\eqref{par}. Using the same definition as in \citet{Chen_al_2012} we obtain the same spectral index of $0.86$ (not shown). 
This anisotropy is consistent with previous measurements in the solar wind that assumed axisymmetry around local mean field \citep{Horbury_al_2008, Podesta_2009, Luo_Wu_2010, Wicks_al_2010, Wicks_al_2011} and supports \citet{GS95} anisotropy that predicts tube-like structures.
This is not surprising given that our strong expansion dataset as well as intervals studied in previous works are characterized by a large value of cross helicity. 
On the contrary it is quite remarkable that GS anisotropy, which is obtained assuming vanishing cross-helicity and axisymmetry, holds in these intervals that do not satisfies these conditions. 
It remains to understand why GS anisotropy emerges when the sampling is in the radial direction, given that non axisymmetric structures are expected when the sampling direction is away from radial one \citep{Verdini_Grappin_2015,Vech_Chen_2016}. 

In the weak-expansion dataset, $0<R<2$, the anisotropy differs substantially from what found in previous works.
The power-law indices of $SF$ are consistent with $1/2,~2/3,~1$ in the perpendicular, displacement, and parallel directions, respectively, provided that the solar wind flow direction is measured accurately (see \citealt{Gerick_al_2017} for effects of uncertainties on the local mean-field direction).
To our knowledge this is the first time that spectral indices matching Boldyrev phenomenology are obtained in solar wind data.

However, a direct measurement of the angle between velocity and magnetic fluctuations fails to obtain the scaling $\theta_\bot\sim\ell_\bot^{1/4}$ that is fundamental in Boldyrev phenomenology. 
Since in this dataset the mean field is expected to be the only symmetry axis, the measured anisotropy indicates that MHD turbulence forms ribbon-like structures, but their origin cannot be attributed to Boldyrev phenomenology.

Another indication that we are observing an anisotropy not consistent with standard (ou pure) Boldyrev theory is the power-law index of the global structure function, which is almost identical in the two datasets with a value close to $2/3$. This is in contrast with the different spectral indices found for the corresponding dominant perpendicular direction in the local analysis (2/3 and 1/2 for the strong- and weak-expansion datasets, respectively).
The different spectral index for the local and global $SF$ in the weak expansion dataset originates from the particular angular distribution of energy, in which an energetically dominant population with 2/3 power-law index appears at oblique angles in the perpendicular plane. 
Such population could be related to compressible fluctuations, which are neglected in the above turbulence phenomenologies. A preliminary attempt to eliminate slow-mode fluctuations does not show any modification with respect to what presented here. Further analysis is required to understand if fast-mode turbulence, which has isotropic energy distribution \citep{Cho_Lazarian_2003,Chandran_2005}, can contribute to the measured anisotropy.

The anisotropy of the weak-expansion dataset is in contrast with that one found in some numerical simulations of homogenous MHD turbulence that support Boldyrev phenomenology \citep{Mason_al_2006,Mason_al_2008,Perez_al_2012,Mallet_al_2016}. Note, however, that the alignement angle, the 3D anisotropy, and the spectral index of global structure functions have not been obtained simultaneously in the above works, while other groups obtain properties that support the GS anisotropy  \citep[e.g.][]{Beresnyak_2014}.
A validation of Boldyrev theory in solar wind data is still missing and would be important for recent theoretical developments which assume its 3D anisotropy\citep{Mallet_al_2015,Chandran_al_2015,Mallet_al_2017a,Mallet_al_2017b,Boldyrev_Louriero_2017,Louriero_Boldyrev_2017}.

It is important to notice that assuming axisymmetry around the mean field would yield an anisotropy consistent with GS theory even for the weak expansion dataset. 
To summarize, our selection criterium allows us to reveal a Boldyrev-like anisotropy which is submerged in GS anisotropy.
How general is this combination of anisotropic structures?
Our selection criterion is supposed to remove expansion effects but we can not prove it to be effective with single-spacecrafts data. In fact, our smallest scale where power-law indices are measured is about $5~\mathrm{Mm}$, above which expansion may still plays a role \citep{Vech_Chen_2016}.
In particular we can not exclude a change of anisotropy for sampling directions different from the radial one. 
In addition, in the weak-expansion dataset, the mean-field axis and the expansion axis are orthogonal. Such configuration produces an asymmetry in the plane perpendicular to the mean field: one of the two field-perpendicular directions is aligned with the radial direction and does not feel the stretching due to expansion, while the other does. Thus, we cannot either exclude that configurations with different angles between the radial and mean-field axis would produce a different combination of Boldyrev and GS anisotropy.

Whether or not expansion is still playing a role in the weak-expansion dataset, our results show that variance anisotropy at large scales and local anisotropy at inertial-range scales are intimately connected.
Such a relation suggests that there is not a unique asymptotic form of turbulent structures but rather a whole variety of structures, originating from a combination of GS and Boldyrev anisotropy (and possibly other ones).

We plan to carry out numerical simulations in the expanding box model to clarify the relation between variance and local anisotropy and to understand if the sampling direction or the angle between the mean field and radial directions may change the Boldyrev-like anisotropy found in the weak-expansion dataset.\\

\textit{Acknowledgments} 
This work has been done within the LABEX PLAS@PAR project, and received
financial state aid managed by the Agence Nationale de la Recherche, as part of
the Programme ``Investissements d'Avenir'' under the reference
ANR-11-IDEX-0004-02.
RG acknowledges support from Programme National Soleil-Terre (PNST/INSU/CNRS).
AV acknowledges L. Matteini and S. Landi for useful discussion on the measure of local anisotropy.
WIND data were obtained from CDAWeb (http://cdaweb.gsfc.nasa.gov).


\end{document}